\def\1ad{\mbox{\normalsize $^1$}}
\def\2ad{\mbox{\normalsize $^2$}}
\def\3ad{\mbox{\normalsize $^3$}}
\def\4ad{\mbox{\normalsize $^4$}}
\def\5ad{\mbox{\normalsize $^5$}}
\def\6ad{\mbox{\normalsize $^6$}}
\def\7ad{\mbox{\normalsize $^7$}}
\def\8ad{\mbox{\normalsize $^8$}}
\def\makefront{\vspace*{1cm}\begin{center}
\def\newtitleline{\\ \vskip 5pt}
{\Large\bf\titleline}\\
\vskip 1truecm
{\large\bf\authors}\\
\vskip 5truemm
\addresses
\end{center}
\vskip 1truecm
{\bf Abstract:}
\abstracttext
\vskip 1truecm}
\documentstyle[12pt]{article}
\def\beq{\begin{equation}}
\def\eeq{\end{equation}}
\def\bea{\begin{eqnarray}}
\def\eea{\end{eqnarray}}
\def\bq{\begin{quote}}
\def\eq{\end{quote}}

\def\gappeq{\mathrel{\rlap {\raise.5ex\hbox{$>$}}
{\lower.5ex\hbox{$\sim$}}}}

\def\lappeq{\mathrel{\rlap{\raise.5ex\hbox{$<$}}
{\lower.5ex\hbox{$\sim$}}}}

\begin{document}
\pagestyle{empty}
\begin{flushright}
{CERN-TH/98-10}
\end{flushright}
\vspace*{5mm}
\begin{center}
{\bf BPS BLACK HOLES, SUPERSYMMETRY AND ORBITS OF EXCEPTIONAL GROUPS} \\
\vspace*{1cm}
{\bf S. Ferrara}$^{*)}$ \\
\vspace{0.3cm}
Theoretical Physics Division, CERN \\
CH - 1211 Geneva 23 \\
\vspace*{2cm}
{\bf ABSTRACT} \\ \end{center}
\vspace*{5mm}
\noindent
We report on duality invariant constraints, which allow a classification of BPS
black holes preserving different fractions of supersymmetry.  We then relate
this analysis to the orbits of the exceptional groups $E_{6(6)}, E_{7(7)}$,
relevant for black holes in five and four dimensions.

\vspace*{2cm}
\begin{center}
{\it Talk given at the Workshop on}\\
{\it ``Quantum Aspects of Gauge Theories, Supersymmetry and Unification"}\\
{\it Neuch\^atel, Switzerland}\\
{\it 18--23 September 1997}
\end{center}
\vspace*{2cm}
\noindent
\rule[.1in]{16.5cm}{.002in}

\noindent
$^{*)}$ Work supported in part by EEC under TMR contract ERBFMRX-CT96-0045 (LNF
Frascati) and by DOE grant DE-FG03-91ER40662.
\vspace*{0.5cm}

\begin{flushleft} CERN-TH/98-10 \\
January 1998
\end{flushleft}
\vfill\eject
%\pagestyle{empty}
%\clearpage\mbox{}\clearpage

\setcounter{page}{1}
\pagestyle{plain}

%INSERT YOUR TEXT HERE
 %%%%% BEGINNING OF NEUCHATEL.TEX %%%%%%%%%%%%%%%
%&latex
%%%%%%%%%%%%%%%%%%%%%%%%%%%%%%%%%%%%%%%%%%%%%%
%                                            %
% Latex file neuchatel.tex                   %
% It needs style file neuchatel.sty.         %
% It may serve as an example for your file.  %
%                                            %
%%%%%%%%%%%%%%%%%%%%%%%%%%%%%%%%%%%%%%%%%%%%%%

%%%%%%%%%%%%%%%%%%%%%%%%%%%%%%%%%%%%%%%%%%%%%%
%                                            %
% Insert now your own Latex definitions.     %
%                                            %
% But do not (!!!) include any definition    %
% concerning pagestyle, margins, length      %
% and width of text. Do not include an own   %
% titlepage or title style.                   %
%                                            %
%%%%%%%%%%%%%%%%%%%%%%%%%%%%%%%%%%%%%%%%%%%%%%
\newcommand{\ft}[2]{{\textstyle\frac{#1}{#2}}}
\newcommand{\QED}{{\hspace*{\fill}\rule{2mm}{2mm}\linebreak}}
\def\dop{{\rm d}\hskip -1pt}
\def\bfone{\relax{\rm 1\kern-.35em 1}}
\def\bfzero{\relax{\rm I\kern-.18em 0}}
\def\inbar{\vrule height1.5ex width.4pt depth0pt}
\def\IC{\relax\,\hbox{$\inbar\kern-.3em{\rm C}$}}
\def\ID{\relax{\rm I\kern-.18em D}}
\def\IF{\relax{\rm I\kern-.18em F}}
\def\IK{\relax{\rm I\kern-.18em K}}
\def\IH{\relax{\rm I\kern-.18em H}}
\def\II{\relax{\rm I\kern-.17em I}}
\def\IN{\relax{\rm I\kern-.18em N}}
\def\IP{\relax{\rm I\kern-.18em P}}
\def\IQ{\relax\,\hbox{$\inbar\kern-.3em{\rm Q}$}}
\def\IR{\relax{\rm I\kern-.18em R}}
\def\IG{\relax\,\hbox{$\inbar\kern-.3em{\rm G}$}}
\font\cmss=cmss10 \font\cmsss=cmss10 at 7pt
\def\ZZ{\relax\ifmmode\mathchoice
{\hbox{\cmss Z\kern-.4em Z}}{\hbox{\cmss Z\kern-.4em Z}}
{\lower.9pt\hbox{\cmsss Z\kern-.4em Z}}
{\lower1.2pt\hbox{\cmsss Z\kern-.4em Z}}\else{\cmss Z\kern-.4em
Z}\fi}
\def\a{\alpha} \def\b{\beta} \def\d{\delta}
\def\e{\epsilon} \def\c{\gamma}
\def\G{\Gamma} \def\l{\lambda}
\def\L{\Lambda} \def\s{\sigma}
\def\cA{{\cal A}} \def\cB{{\cal B}}
\def\cC{{\cal C}} \def\cD{{\cal D}}
\def\cF{{\cal F}} \def\cG{{\cal G}}
\def\cH{{\cal H}} \def\cI{{\cal I}}
\def\cJ{{\cal J}} \def\cK{{\cal K}}
\def\cL{{\cal L}} \def\cM{{\cal M}}
\def\cN{{\cal N}} \def\cO{{\cal O}}
\def\cP{{\cal P}} \def\cQ{{\cal Q}}
\def\cR{{\cal R}} \def\cV{{\cal V}}\def\cW{{\cal W}}
%
%%%%%%%%%%%%%%%%%%%%%%%%%%%%%%%%%%%%%%%%%%%%%%%%%%%%%%%%%%%%%%
%%%%% misc macros %%%%%
%%%%%%%%%%%%%%%%%%%%%%%%%%%%%%%%%%%%%%%%%%%%%%%%%%%%%%%%%%%%%%
%
\def\crr{\crcr\noalign{\vskip {8.3333pt}}}
\def\tilde{\widetilde}
\def\bar{\overline}
\def\us#1{\underline{#1}}
\let\shat=\hat
\def\hat{\widehat}
\def\hyp{\vrule height 2.3pt width 2.5pt depth -1.5pt}
\def\square{\mbox{.08}{.08}}
\def\Coeff#1#2{{#1\over #2}}
\def\Coe#1.#2.{{#1\over #2}}
\def\coeff#1#2{\relax{\textstyle {#1 \over #2}}\displaystyle}
\def\coe#1.#2.{\relax{\textstyle {#1 \over #2}}\displaystyle}
\def\half{{1 \over 2}}
\def\shalf{\relax{\textstyle {1 \over 2}}\displaystyle}
\def\dag#1{#1\!\!\!/\,\,\,}
\def\to{\rightarrow}
\def\notin{\hbox{{$\in$}\kern-.51em\hbox{/}}}
\def\shdot{\!\cdot\!}
\def\ket#1{\,\big|\,#1\,\big>\,}
\def\bra#1{\,\big<\,#1\,\big|\,}
\def\equaltop#1{\mathrel{\mathop=^{#1}}}
\def\Trbel#1{\mathop{{\rm Tr}}_{#1}}
\def\inserteq#1{\noalign{\vskip-.2truecm\hbox{#1\hfil}
\vskip-.2cm}}
\def\attac#1{\Bigl\vert
{\phantom{X}\atop{{\rm\scriptstyle #1}}\phantom{X}}}
\def\exx#1{e^{{\displaystyle #1}}}
\def\del{\partial}
\def\delbar{\bar\partial}
\def\nex#1{$N\!=\!#1$}
\def\dex#1{$d\!=\!#1$}
\def\cex#1{$c\!=\!#1$}
\def\eg{{\it e.g.}} \def\ie{{\it i.e.}}
%\catcode`\@=12

%%%%%%%%%%%%%%%%%%%%%%%%%%%%%%%%%%%%%%%%%%%%%%%%%%%%%%%%%%%%%%
%%%%%%%%%%%%%%%%%%%%%%%%%%%%%%%%%%%%%%%%%%%%%%%%%%%%%%%%%%%%
%\draft
%%%%%%%%%%%% macros and references %%%%%%%%%%%%%%%%%%%%%%%%%
%
\def\cS{{\cal K}}
\def\IE{\relax{{\rm I\kern-.18em E}}}
\def\IGam{\relax{{\rm I}\kern-.18em \Gamma}}
\def\IGa{\IA}
\def\IA{\relax{\hbox{{\rm A}\kern-.82em {\rm A}}}}
\let\picfuc=\fp
\def\hata{{\shat\a}}
\def\hatb{{\shat\b}}
\def\hatA{{\shat A}}
\def\hatB{{\shat B}}
\def\bv{{\bf V}}
%

%%%%%%%%%%%%%%%%%%%%%%%%%%%%%%%%%%%%%%%%%

\def\titleline{
%%%%%%%%%%%%%%%%%%%%%%%%%%%%%%%%%%%%%%%%%%%%%%
%                                            %
% Insert now the text of your title.         %
% Make a linebreak in the title with         %
%                                            %
%            \newtitleline                   %
%                                            %
%%%%%%%%%%%%%%%%%%%%%%%%%%%%%%%%%%%%%%%%%%%%%%
BPS BLACK HOLES, SUPERSYMMETRY
\newtitleline
AND ORBITS OF EXCEPTIONAL GROUPS
%%%%%%%%%%%%%%%%%%%%%%%%%%%%%%%%%%%%%%%%%%%%%%
}
\def\authors{
%%%%%%%%%%%%%%%%%%%%%%%%%%%%%%%%%%%%%%%%%%%%%%
%                                            %
%  Insert now the name (names) of the author %
%  (authors).                                %
%  In the case of several authors with       %
%  different addresses use e.g.              %
%                                            %
%             \1ad  , \2ad  etc.             %
%                                            %
%  to indicate that a given author has the   %
%  address number 1 , 2 , etc.               %
%                                            %
%%%%%%%%%%%%%%%%%%%%%%%%%%%%%%%%%%%%%%%%%%%%%%
Sergio FERRARA  %\1ad , Second Author \2ad
%%%%%%%%%%%%%%%%%%%%%%%%%%%%%%%%%%%%%%%%%%%%%
}
\def\addresses{
%%%%%%%%%%%%%%%%%%%%%%%%%%%%%%%%%%%%%%%%%%%%%%
%                                            %
% List now the address. In the case of       %
% several addresses list them by numbers     %
% using e.g.                                 %
%                                            %
%             \1ad , \2ad   etc.             %
%                                            %
% to numerate address 1 , 2 , etc.           %
%                                            %
%%%%%%%%%%%%%%%%%%%%%%%%%%%%%%%%%%%%%%%%%%%%%%
%\1ad
Theoretical Physics Division, CERN, CH-1211 Geneva 23\\
%\2ad
%Institute of Physics, University, \\
%A-Street 1, A-Town
%%%%%%%%%%%%%%%%%%%%%%%%%%%%%%%%%%%%%%%%%%%%%%%
}
\def\abstracttext{
%%%%%%%%%%%%%%%%%%%%%%%%%%%%%%%%%%%%%%%%%%%%%%%
%                                             %
% Insert now the text of your abstract.       %
%                                             %
%%%%%%%%%%%%%%%%%%%%%%%%%%%%%%%%%%%%%%%%%%%%%%%
We report on duality invariant constraints, which allow a classification of BPS
black holes preserving different fractions of supersymmetry.  We then relate
this analysis to the orbits of the exceptional groups $E_{6(6)}, E_{7(7)}$,
relevant for black holes in five and four dimensions.}
%%%%%%%%%%%%%%%%%%%%%%%%%%%%%%%%%%%%%%%%%%%%%%%%%%%%%%%%%%%%%%%
\makefront
%%%%%%%%%%%%%%%%%%%%%%%%%%%%%%%%%%%%%%%%%%%%%%%%
%                                              %
%  Insert now the remaining parts of           %
%  your article.                               %
%                                              %
%%%%%%%%%%%%%%%%%%%%%%%%%%%%%%%%%%%%%%%%%%%%%%%%
\section{Introduction}
Impressive results have recently been obtained in the study of general properties of BPS black
holes in supersymmetric theories of gravity.
The latter are described by
 string theory and M-theory \cite{string} whose symmetry properties are encoded
in extended supergravity effective field theories.

Of particular interest are extremal black holes in four and five dimensions which
correspond to BPS saturated states \cite{black} and whose ADM mass depends,
 beyond the
quantized values of electric and magnetic charges, on the asymptotic value
 of scalars at infinity.
The latter describe the moduli space of the theory
Another physical relevant quantity, which depends only on quantized electric
 and magnetic charges,
 is the black hole entropy,
which can be defined macroscopically, through the Bekenstein-Hawking
 area-entropy relation
or microscopically, through D-branes techniques \cite{dbr} by counting
of microstates \cite{micros}.
It has been further realized that the scalar fields, independently of
their values
at infinity, flow towards the black hole horizon to a fixed value of pure
 topological
nature given by a certain ratio of electric and magnetic charges \cite{fks}.
These ``fixed scalars'' correspond to the extrema of the ADM mass
in moduli space while the black-hole entropy  is the  value of the
 squared
ADM mass at this point in $D=4$  \cite{feka1} \cite{feka2} and the power $3/2$
of the ADM mass in $D=5$.
In four dimensional theories with $N>2$, extremal black-holes preserving one supersymmetry
have the further property that all central charge eigenvalues other than the one
equal to the BPS mass flow to zero for ``fixed scalars''.

The entropy formula turns out to be in all cases a U-duality invariant
expression
(homogeneous of degree two in $D=4$ and of degree $3/2$ in $D=5$)
built out of electric and magnetic charges and as such
can be in fact also computed through certain (moduli-independent) topological
quantities which only depend on the nature of the U-duality groups and the
appropriate representations
of electric and magnetic charges.
More specifically, in the $N=8$, $D=4$ and $D = 5$ theories  the entropy was shown to correspond  to the
unique quartic $E_7$ and cubic $E_6$ invariants built with the 56 and 27 dimensional
representations respectively
\cite{kall}, \cite{feka1}.
In this report  we respectively describe the invariant classification of BPS states preserving
different numbers of supersymmetries in $D = 4$ and 5 and then relate this analysis
to the theory of BPS orbits of the exceptional groups $E_{7(7)}$ and $E_{6(6)}$.

%%%%%%%%%%%%%%%%%%%%%%%%%%%%%%%%%%%%%%%%%%%%%%%%%%%%%%%%%%%%%%%%%%%%%%%

\section{BPS Conditions for Enhanced Supersymmetry}
In this section we will describe U-duality invariant constraints
on the multiplets of quantized charges in the case of BPS black holes
 whose background preserves more than one supersymmetry \cite{fema}.

We will still restrict our analysis to four and five dimensional cases
for which three possible cases exist {\it i.e.} solutions preserving
 $1/8$, $1/4$ and $1/2$ of the original  supersymmetry (32 charges).

The invariants may only be non zero on solutions preserving $1/8$ supersymmetry.
In dimensions $6\le D \le 9$ black holes may only preserve $1/4$ or $1/2$
supersymmetry, and no associated invariants exist in these cases.

The description which follows also make contact with the D-brane
 microscopic calculation, as it will appear
obvious from the formulae given below.
We will first consider the five dimensional case.

In this case, BPS states preserving $1/4$  of supersymmetry correspond
 to the invariant constraint $I_3 (27) =0$ where $I_3$ is the $E_6$ cubic invariant
  \cite{feka1}.
This corresponds to the $E_6$ invariant statement that the {\bf 27} is a null vector
with respect to the cubic norm.
As we will show in a moment, when this condition is fulfilled
 it may be shown that two of the central charge eigenvalues
are equal in modulus.
The generic configuration has 26 independent charges.

Black holes corresponding to $1/2$ BPS states correspond to null vectors
which are critical, namely
\begin{equation}
\partial I(27) =0
\end{equation}

In this case the three central charge eigenvalues are equal in modulus
and a generic charge vector has 17 independent components.

To prove the above statements, it is useful to compute the cubic invariant
in the normal frame, given  by:
\begin{eqnarray}
I_3(27) &=& Tr(Z\IC )^3 \nonumber\\
&=& 6 (e_1+e_2)(e_1+e_3)(e_2+e_3)\nonumber\\
& = & 6 s_1s_2s_3
\end{eqnarray}
where:
\begin{eqnarray}
e_1 &=& {1 \over 2} (s_1 + s_2 - s_3 ) \nonumber\\
e_2 &=& {1 \over 2} (s_1 - s_2 + s_3 ) \nonumber\\
e_3 &=& {1 \over 2} (- s_1 + s_2 + s_3 )
\end{eqnarray}
are the eigenvalues of the traceless antisymmetric $8\times 8$
matrix.
We then see that if $s_1=0$ then $\vert e_1\vert = \vert e_2\vert $,
and if $s_1= s_2 =0$ then $\vert e_1\vert = \vert e_2\vert = \vert e_3\vert $.
To count the independent charges we must add to the eigenvalues
the angles given by $USp(8)$ rotations.
The subgroup of $USp(8)$ leaving two eigenvalues invariant is $USp(2)^4$,
which is twelve dimensional.
The subgroup of $USp(8)$ leaving invariant one eigenvalue is $USp(4) \times USp(4)$,
which is twenty dimensional.
The angles are therefore $36-12=24$ in the first case, and $36-20=16$ in the second case.
This gives rise to configurations with 26 and 17 charges respectively, as promised.

Taking the case of Type II on $T^5$ we can choose $s_1$ to correspond
to a solitonic five-brane charge, $s_2$ to a fundamental string
winding charge along some direction and $s_3$ to Kaluza-Klein
momentum along the same direction.

The basis chosen in the  above example is S-dual to the D-brane basis
usually chosen for describing black holes in Type IIB on $T_5$.
All other bases are related by U-duality to this particular choice.
We also observe that the above analysis relates the cubic invariant to the picture  of
intersecting branes since a three-charge $1/8$ BPS configuration with non vanishing entropy
  can be thought as obtained by intersecting three single charge $1/2$ BPS configurations
\cite{bdl}, \cite{bll}, \cite{lptx}

By using the S--T-duality decomposition we see that the cubic
invariant reduces to $I_3(27)=10_{-2} 10_{-2} 1_4 + 16_1 16_1
10_{-2}$.
The 16 correspond to D-brane charges, the 10 correspond to the 5 KK
directions and winding of wrapped fundamental strings, the 1
correspond to the N-S five-brane charge.

We see that to have a non-vanishing area we need a configuration with
three non-vanishing N-S charges or two D-brane charges and one N-S charge.

Unlike the 4-$D$ case, it is impossible to have a non-vanishing entropy
for a configuration only carrying D-brane charges.

We now turn to the four dimensional case.

In this case the situation is more subtle because the condition for the 56
to be a null vector (with respect to the quartic norm) is not sufficient to enhance the supersymmetry.
This can be easily seen by going to the normal frame where it can be shown
that for a null vector there are not, in general,
coinciding eigenvalues.
The condition for $1/4$ supersymmetry is that the gradient of the quartic invariant vanish.

The invariant condition for $1/2$ supersymmetry is that the second
 derivative projected into the adjoint representation
of $E_7$ vanish.
This means that, in the symmetric quadratic polynomials of second derivatives,
only terms in the 1463 of $E_7$ are non-zero.
Indeed, it can be shown, going to the normal frame for the 56 written as a skew
$8 \times 8$ matrix, that the above conditions imply two and four eigenvalues
 being equal respectively.

The independent charges of $1/4$ and $1/2$ preserving supersymmetry
are 45 and 28 respectively.

To prove the latter assertion,
it is sufficient to see that the two charges normal-form matrix is left
 invariant by $USp(4) \times USp(4)$,
while the one charge matrix is left invariant by $USp(8)$ so the $SU(8)$ angles are
$63-20=43$ and $63-36=27$ respectively.

The generic $1/8$ supersymmetry preserving  configuration of the 56 of $E_7$
with non vanishing
entropy has five independent parameters in the normal frame and $51=63-12$
$SU(8)$ angles.
This is because the compact little group of the normal frame is $SU(2)^4$.
The five parameters describe the four eigenvalues and an overall phase
of an $8\times 8$ skew diagonal matrix.

If we allow the phase to vanish, the 56 quartic norm just simplifies
as in the five dimensional case:
\begin{eqnarray}
 I_4(56) &= &s_1s_2s_3s_4
= (e_1+e_2+e_3+e_4)(e_1+e_2-e_3-e_4)\nonumber\\
&\times& (e_1-e_2-e_3+e_4)(e_1-e_2+e_3-e_4)
\label{i4inv}
\end{eqnarray}
where $e_i$ ($i=1,\cdots, 4$) are the four eigenvalues.

$1/4$ BPS states correspond to $s_3=s_4=0$ while $1/2$ BPS states correspond to
$s_2=s_3=s_4=0$.

An example of this would be a set of four D-branes oriented along 456, 678, 894, 579
(where the order of the three numbers indicates the orientation of the brane).
Note that in choosing the basis the sign of the D-3-brane charges is important; here they are chosen
such that taken together with positive coefficients they form a BPS object.
The first two possibilities  ($I_4 \neq 0$ and $I_4 =0$,
 ${\partial I_4 \over \partial q^i \neq 0}$) preserve $1/8$ of the supersymmetries, the
third (${\partial I_4 \over \partial q^i} = 0$, ${\partial^2 I_4 \over \partial q^i\partial q^j}\vert_{Adj~E_7} \neq 0$)
$1/4$ and the last (${\partial^2 I_4 \over \partial q^i\partial q^j}\vert_{Adj~E_7} = 0$) $1/2$.

It is interesting that there are two types of $1/8$ BPS solutions.
In the supergravity description, the difference between them is that the first case has non-zero
horizon area.
If $I_4 <0$ the solution is not BPS.
This case corresponds, for example, to changing the sign of one of the three-brane
charges discussed above.
By U-duality transformations we can relate this to configurations of branes at angles such as in \cite{balale}

Going from four to five dimensions it is natural to decompose the $E_7 \to E_6 \times O(1,1)$
where $E_6$ is the duality group in five dimensions and $O(1,1)$ is
 the extra T-duality that appears when we compactify from five to four
dimensions.
According to this decomposition, the representation breaks as: ${\bf 56} \to {\bf 27}_1 + {\bf 1}_{-3}
+ {\bf 27}'_{-1} + {\bf 1}_3$ and the quartic invariant becomes:
\begin{equation}
  \label{decomp}
  {\bf 56}^4 = ({\bf 27}_1)^3{\bf 1}_{-3}+({\bf 27^\prime}_{-1})^3{\bf 1}_{3} +
  {\bf 1}_{3} {\bf 1}_{3}{\bf 1}_{-3}{\bf 1}_{-3}
+ {\bf 27}_1{\bf 27}_1{\bf 27}'_{-1}{\bf 27}'_{-1} +{\bf 27}_1{\bf 27}'_{-1} {\bf 1}_{3}{\bf 1}_{-3}
\end{equation}
The ${\bf 27}$ comes from point-like charges in five dimensions and the ${\bf 27}'$
comes from string-like charges.

Decomposing the U-duality group into T- and S-duality groups, $E_7 \to SL(2,\IR)\times O(6,6)$,
we find ${\bf 56} \to ({\bf 2,12}) + ({\bf 1,32})$ where the first term
corresponds to N-S charges and the second term to D-brane charges.
Under this decomposition the quartic invariant (\ref{i4inv}) becomes
${\bf 56}^4 \to {\bf 32}^4 + ({\bf 12 . 12'})^2 + {\bf 32}^2.{\bf 12 . 12'}$.
This means that we can have configurations with a non-zero area that carry
only D-brane charges, or only N-S charges or both D-brane and N-S charges.

It is remarkable that $E_{7(7)}$-duality gives additional restrictions on the BPS states
other than the ones merely implied by the supersymmetry algebra.
The analysis of double extremal black holes implies that $I_4$ be semi-definite positive
for BPS states.
>From this fact it follows that configurations preserving $1/4$ of supersymmetry must have eigenvalues
equal in pairs, while configuratons with three coinciding eigenvalues are not BPS.

To see this,  it is sufficient to write the quartic invariant in the normal frame
basis.
A generic skew diagonal $8\times 8$ matrix depends on four complex eigenvalues $z_i$.
These eight real parameters can be understood using the decomposition \cite{trig}:
\begin{equation}
  {\bf 56} \to  ( {\bf 8_v,2,1,1}) + ({\bf 8_s,1,2,1})+ ({\bf 8_c,1,1,2})
 + ({\bf 1,2,2,2})
\end{equation}
under
\begin{equation}
  E_{7(7)} \to O(4,4) \times SL(2,\IR)^3
\end{equation}
Here $O(4,4)$ is the little group of the normal form and the
 $({\bf 2,2,2})$ are the four complex skew-diagonal elements.
We can further use $U(1)^3 \subset SL(2,\IR)^3$ to further remove three relative phases
so we get the five parameters $z_i = \rho_i e^{{\rm i} \phi/4}$ ($i=1,\cdots, 4$).

The quartic invariant, which is also the unique $SL(2,\IR)^3$ invariant built with the $({\bf 2,2,2})$,
becomes \cite{fema}:
\begin{eqnarray}
 I_4&= &\sum_i \vert z_i \vert^4 -2 \sum_{i<j} \vert z_i \vert^2 \vert z_j \vert ^2
 + 4 (z_1z_2z_3z_4 + {\bar z_1}{\bar z_2}{\bar z_3}{\bar z_4} )\nonumber\\
&= &(\rho_1 + \rho_2 + \rho_3 + \rho_4) (\rho_1 + \rho_2 - \rho_3 - \rho_4)
\times
(\rho_1 - \rho_2 + \rho_3 + \rho_4)(\rho_1 - \rho_2 - \rho_3 + \rho_4)\nonumber\\
&+& 8\rho_1\rho_2\rho_3\rho_4(cos \phi -1)
\end{eqnarray}

The last term is semi-definite negative.
The first term, for $\rho_1=\rho_2=\rho$ becomes:
\begin{equation}
  -[4\rho^2 -(\rho_3+\rho_4)^2](\rho_3-\rho_4)^2
\end{equation}
which is negative unless $\rho_3=\rho_4$.
So $1/4$ BPS states must have
\begin{equation}
\rho_1=\rho_2 > \rho_3=\rho_4~,\quad cos \phi =1
\end{equation}
For $\rho_1=\rho_2 =\rho_3=\rho$, the first term in $I_4$ becomes:
\begin{equation}
  -(3\rho + \rho_4)(\rho - \rho_4)^3
\end{equation}
so we must also have $\rho_4=\rho~,\quad cos \phi=1$
which is the $1/2$ BPS condition.

An interesting case, where $I_4$ is negative, corresponds to a configuration carrying
electric and magnetic charges under the same gauge group, for example a 0-brane
plus 6-brane configuration which is dual to a K--K-monopole plus K--K-momentum
 \cite{ko}, \cite{shein}.
This case corresponds to $z_i=\rho e^{{\rm i} \phi/4}$ and the phase is $tan \phi/4 =e/g$
where $e$ is the electric charge and $g$ is the magnetic charge.
Using (\ref{i4inv}) we find that $I_4<0$ unless the solution is purely electric or
purely magnetic.
In \cite{pol} it was suggested that  $0+6$ does not form a supersymmetric state.
Actually, it was shown in \cite{tay} that a $0+6$ configuration can be
 T-dualized into a non-BPS configuration of four intersecting D-3-branes.
Of course, $I_4$ is negative for both configurations.
Notice that even though these two charges are Dirac dual (and U-dual)
they are not S-dual in the sense of filling out an $SL(2,\ZZ)$
multiplet.
In fact, the K--K-monopole forms an $SL(2,\ZZ)$ multiplet with a fundamental string winding
charge under S-duality \cite{sen}

%%%%%%%%%%%%%%%%%%%%%%%%%%%%%%%%%%%%%%%%%%%%%%%%%%%%%%%%%%%%%%%%%%%%%%%%%
%%%%%%%%%%%%%%%%%%%%%%%%%%%%%%%%%%%%%%%%%%%%%%%%%%%%%%%%%%%%%%%%%%%%%%%
\section{Duality Orbits fo BPS States Preserving Different Numbers of Supersymmetries}
In this last section we give an invariant classification of BPS black holes preserving
 different numbers of supersymmetries in terms
of orbits of the {\bf 27} and the {\bf 56} fundamental representations of the duality
groups $E_{6(6)}$ and $E_{7(7)}$ resperctively \cite{fg}, \cite{stelle}.

In five dimensions the generic orbits preserving $1/8$ supersymmetry correspond to the
26 dimensional orbits $E_{6(6)}/F_{4(4)}$
so we may think the generic 27 vector of $E_6$ parametrized by a point in this orbit
and its cubic norm (which actually equals the square of the black-hole entropy).

The light-like orbit, preserving $1/4$ supersymmetry, is the 26
dimensional coset $E_{6(6)}/O(5,4) \odot T_{16}$
where $ \odot $ denotes the semidirect product.

The critical orbit, preserving maximal $1/2$ supersymmetry (this correspond to
${\partial I_4 \over \partial q^i }\neq 0$) correspond to the 17 dimensional space
\begin{equation}
  {E_{6(6)} \over O(5,5) \odot T_{16}}
\end{equation}
In the four dimensional case, we have two inequivalent 55 dimensional orbits
corresponding to the cosets $E_{7(7)}/E_{6(2)}$ and $E_{7(7)}/E_{6(6)}$ depending
on whether $I_4(56)>0$ or $I_4(56)<0$.
The first case corresponds to $1/8$ BPS states whith non-vanishing entropy,
while the latter corresponds to non BPS states.

There is an additional 55 dimensional light-like orbit ($I_4 =0$) preserving
$1/8$ supersymmetry given by ${E_{7(7)} \over F_{4(4)}\odot T_{26}}$.

The critical light-like orbit, preserving $1/4$ supersymmetry, is the
45 dimesnsional coset  $E_{7(7)}/O(6,5) \odot (T_{32}\oplus T_1)$

The critical orbit corresponding to maximal $1/2$ supersymmetry is described
by the 28 dimensional quotient space
\begin{equation}
{E_{7(7)} \over E_{6(6)}\odot T_{27}}
\end{equation}
We actually see that the counting of parameters in terms of invariant
orbits reproduces the counting previously made in terms
of normal frame parameters and angles.
The above analysis makes a close parallel between BPS states preserving different numbers of
supersymmetries with time-like, space-like and light-like vectors in Minkowski space.

%%%%%%%%%%%%%%%%%%%%%%%%%%%%%%%%%%%%%%%%%%%%%%%%%%%%%%%%%%%%%%%%%%%%%%%%%
%%%%%%%%%%%%%%%%%%%%%%%%%%%%%%%%%%%%%%%%%%%%%%%%%%%%%%%%%%%%%%%%%%%%%%%
\section*{Acknowledgements}
The results of this report have been obtained  in collaborations with J.M. Maldacena and M. Gunaydin.

%%%%%%%%%%%%%%%%%%%%%%%%%%%%%%%%%%%%%%%%%%%%%%%%%%%%%%%%%%%%%%%%%%%%%%%%%
%%%%%%%%%%%%%%%%%%%%%%%%%%%%%%%%%%%%%%%%%%%%%%%%%%%%%%%%%%%%%%%%%%%%%%%

\end{document}